\documentclass{iau}
\usepackage{graphicx}

\title[Dusty origin of the BLR in AGN] %% give here short title %%
{Dusty origin of the Broad Line Region in active galaxies\footnote{based on observations made with the Southern African Large Telescope (SALT)}}

\author[B.\ Czerny et al.]   %% give here short author list %%
{Bozena Czerny, Krzysztof Hryniewicz, Janusz Kaluzny, Ishita Maity, Wojtek Pych}

\affiliation{Copernicus Astronomical Center, Bartycka 18, 00-716 Warsaw, Poland  \\ email: {\tt bcz@camk.edu.pl,krhr@camk.edu.pl }}

\pubyear{2012}
\volume{290}  %% insert here IAU Symposium No.
\jname{Feeding compact objects: Accretion on all scales}
\editors{C.\ M. Zhang, T. Belloni, M.\ M\'endez \& S.\ N.\ Zhang, eds.}

\begin{document}

\maketitle

\begin{abstract}
The most characteristic property of active galaxies, including quasars, are
prominent broad emission lines. I will discuss an interesting possibility that 
dust is responsible for this phenomenon. The dust is known to be present in quasars 
in the form of a dusty/molecular torus which results in complexity of the appearance
 of active galaxies. However, this dust is located further from the black hole than 
the Broad Line Region.  We propose that the dust is present also closer in and 
it is actually responsible for formation of the broad emission lines. The argument 
is based on determination of the temperature of the disk atmosphere underlying the
 Broad Line Region: it is close to 1000 K, independently from the black hole mass 
and accretion rate of the object. The mechanism is simple and universal but leads 
to a considerable complexity of the active nucleus surrounding. The understanding
the formation of BLR opens a way to use it reliably - in combination with 
reverberation measurement of its size - as standard candles in cosmology.
\keywords{Accretion, accretion disks; Black hole physics; Cosmology}
\end{abstract}

\firstsection % if your document starts with a section,
\section{Introduction}

Increasing spatial resolution allows to observe directly parts of active galactic nuclei. However,
the innermost part is still unresolved and the information about its geometrical structure comes 
from spectroscopy. This structure is very complex, so vigorous discussions of inflow/outflow 
are still going on. Among this unsolved problems is the origin of Broad Emission Lines. 

Broad Emission Lines come from a specific Broad Emission Line Region (BLR), as seen in well separation of 
the Broad and narrow components in the emission lines of active galaxies. They cover significant fraction 
of the AGN sky, between 10 and 30 \%, as implied by their luminosity, but they are rarely seen in
absorption (for rare exceptions, see e.g. Risaliti et al. 2011) which argues for the flat 
configuration (e.g. Nikolajuk et al. 2005, 
Collin et al. 2006), mostly hidden inside 
the outer dusty-molecular torus. The motion of the material is roughly consistent with 
circular Keplerian orbits (Wandel et al. 1999), but with significant additional velocity dispersion. Careful studies
indicate two components of the BLR (Collin-Souffrin et al. 1988): High Ionization Lines (HIL; e.g. CIV) 
and Low Ionization Lines (LIL; e.g. H$\beta$, Mg II). HIL show blueshift with respect to narrow lines and they
are considered to be comming from outflowing wind. LIL do not show significant shift so cannot come from outflow,
but the region cannot be supported by the hydrostatic equilibrium. Therefore, the dynamics of LIL is
difficult to understand from theoretical point of view, and it is also not resolved by direct
monitoring (e.g. Peterson et al. 2004). 

\begin{figure}[t]
% \vspace*{-2.0 cm}
\begin{center}
\hfill~
 \includegraphics[width=0.6\textwidth]{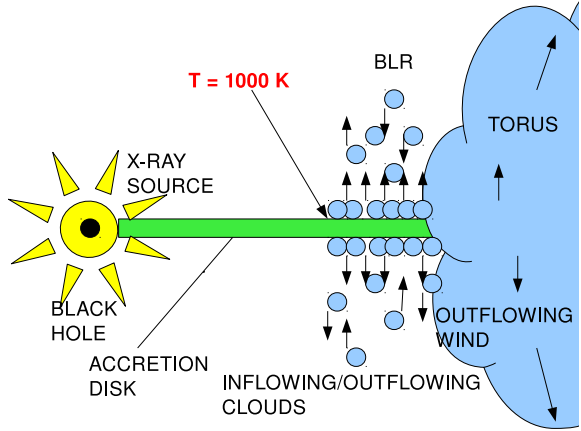}
\hfill~
% \includegraphics[width=0.38\textwidth]{best_both.eps}
%\hfill~
% \vspace*{-1.0 cm}
 \caption{Schematic picture of the formation of the BLR. The outflow starts where the disk effective temperature drops below 1000 K and dust formation can proceed. Dusty outflow changes into a failed wind above the disk, where the irradiation leads to dust evaporation, the radiation pressure suport rapidly decreases and the material falls back onto the disk.}
\end{center}
\end{figure}

\section{Accretion Disk Temperature at the Onset of BLR}

Here we show that the combination of the simplest accretion disk theory and the reverberation
measurement of the BLR distance leads to surprising conclusion: the effective temperature of the
accretion disk at the radius measured from reverberation is universal and equal to 1000 K. The argument goes
in the following way (for more details, see Czerny \& Hryniewicz 2011).

The measurements of the delay of the H$\beta$ line with respect to the continuum at 5100 A for close to 40 objects
allowed to determine the relation between the BLR size, as measured from such delay, and the monochromatic flux
(Kaspi et al. 2000, Peterson et al. 2004). The most recent form of this relation, corrected for the starlight 
contamination, was derived by Bentz et al. (2009). This relation can be written in the following way
\begin{equation}
\log R_{BLR}[{\rm H}\beta] = 1.538 \pm 0.027 + 0.5 \log L_{44,5100},
\label{Bentz}
\end{equation}
where $R_{BLR}[{\rm H}\beta]$ is in light days and $L_{44,5100}$ is the monochromatic luminosity at 
5100 \AA~ measured
in units of $10^{44}$ erg s$^{-1}$. The value 0.027 is the error in the best-fit vertical 
normalization.

Now, from the simplest theory of the alpha disk (Shakura \& Sunyaev 1973) we can find the monochromatic flux. The
asymptotic formula is usually known in the form
\begin{equation}
L_{\nu} \propto \nu^{1/3},
\end{equation}
but the proportionality coefficient is also known, and depends on the black hole mass, $M$, and accretion rate,
$\dot M$ as $(M \dot M)^{2/3}$. Taking into account this dependence, physical constants and fixing the frequency at
the value corresponding to the wavelength 5100 A, we have
\begin{equation}
\log L_{44,5100} = {2 \over 3} \log (M \dot M) - 43.8820 + \log \cos i.
\label{L5100}
\end{equation} 
This formula allows us to calculate the BLR radius from the Eq.~\ref{Bentz}, if the
product of ($M \dot M$) is known.

The Shakura-Sunyaev accretion disk theory allows us to calculate the effective temperature of the disk at that 
radius as a function of the black hole mass and accretion rate
\begin{equation}
T_{eff} = \left ({3GM \dot M \over 8 \pi r^3 \sigma_B}\right )^{0.25},\label{Teff}
\end{equation}
where $\sigma_B$ is the Stefan-Boltzmann constant. Since BLR is far from the central black hole, the inner boundary effect can be neglected.

When we combine Eqs.~\ref{Bentz}, \ref{L5100}, and \ref{Teff}, the dependence on the unknown mass and accretion rate vanishes, and we obtain a single value for all the sources in the sample
\begin{equation}
T_{eff} = 995  \pm 74 {\rm K},
\end{equation}
where the error reflects the error in the constant in Eq.~\ref{Bentz}. We stress that the resulting value does not depend either on the black hole mass or accretion rate. The value is universal, for all sources.

\section{BLR Formation Mechanism}

The value of the universal temperature of $\sim 1000$ K immediately hints that the physical mechanism of the BLR
formation is related to the dust formation in the accretion disk atmosphere. The dust opacity is large, and even 
moderate radiation flux from the underlying disk can push the dusty material off the disk surface. This dusty 
outflow, however, cannot proceed too far from the disk plane. Lifted material is irradiated by the strong 
UV and X-ray flux from the central region, the material heats up, the dust evaporates, the opacity rapidly decreases,
and in the absence of the strong radiation pressure the material fall back again onto the disk surface. Therefore,
such a failed wind can account for lifting the material relatively high above the disk plane and at the same time
does not give a strong systematic signatures of outflow since both outflow and inflow are present.

The overall geometry is shown in Fig.~1. The BLR region starts were the disk effective temperature, without 
irradiation from the central region, falls below 
1000 K, and it ends up at larger distance when the material has the temperature below 1000 K even if the 
irradiation is included. This outer radius marks the transition to the outer dusty/molecular torus, where the 
irradiation is too weak to 
lead to dust evaporation and the outflow proceeds.

\section{Application to the Dark Energy Study}

It was already proposed by Watson et al (2011) that Eq.~2.1 can be used to determine the distance to a quasar in a way independent from the redshift. Reverberation measures the BLR size directly from the time delay, and Eq.~2.1 gives us the intrinsic monochromatic luminosity. This, combined with the observed monochromatic luminosity, gives the luminosity distance to the source. If a number of distant quasars are measured, a luminosity distance vs. redshift diagram can be constructed. Our interpretation of the BLR formation supports the view that Eq.~2.1 can apply to high redshift objects as well, since the mechanism is based on dust formation, this in turn is connected with medium metalicity, and the metalicity of distant quasars is known to be roughly solar.

With this in mind, we started a monitoring program with the Southern Africal Large Telescope. We choose intermediate quasars and Mg II line which belongs to LIL class, as H$\beta$, and should not show rapid intrinsic variations characteristic for CIV line.

\section{Spectroscopy}

\subsection{Observations and Data Reduction}

In order to test the feasibility of spectroscopic studies of the corresponding
objects we have obtained a pair of medium resolution spectra of the quasar
LBQS 2113-4538.
The observations have been performed using the Robert Stobie Spectrograph
on Southern Africal Large Telescope in the service mode. The observing data
was collected in two blocks, in the nights May 15/16, 2012 (UT) and July 30/31, 2012 (UT).
Each block consists of a pair of 978 second exposures of the target spectrum
followed by an exposure of the spectrum of the Argon calibration lamp, and
a set of flat-field images.

Initial data reduction steps: gain correction, cross-talk correction,
overscan bias subtraction and amplifier mosaicking have been performed
by the SALT Observatory staff using a semi-automated pipeline
from the SALT PyRAF package.

Flat-field correction and further reduction steps have been performed
using procedures within the IRAF 
package\footnote{IRAF is distributed by the National Optical Astronomy
Observatories, which are operated by the Association of Universities
for Research in Astronomy, Inc., under cooperative agreement with the
NSF.}.
The pairs of adjacent spectrum images of LBQS 2113-4538 have been
combined into a single image. This enabled us to efficiently reject
cosmic rays and raise the signal to noise ratio.

Identification of the lines in the calibration lamp spectrum, wavelength
calibration, image rectification and extraction of one dimensional spectra
have been done using functions from noao.twodspec package within IRAF.

\subsection{Preliminary results}

\begin{figure}[t]
% \vspace*{-2.0 cm}
\begin{center}
\hfill~
 \includegraphics[width=0.4\textwidth]{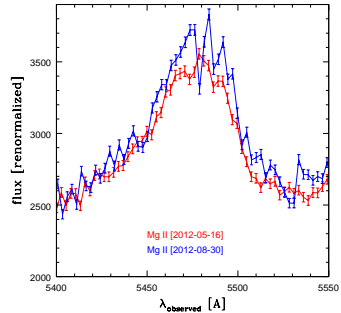}
\hfill~
% \includegraphics[width=0.38\textwidth]{best_both.eps}
%\hfill~
% \vspace*{-1.0 cm}
 \caption{Two observations of the Mg II line of the quasar LBQS 2113-4538, separated by $\sim 100$ days (preliminary results) made with SALT. The shape of the line has clearly changed, and the line intensity has changed by 9 \%, which allows for a systematic monitoring of this source.}
\end{center}
\end{figure}

In Fig. 2 we present the two profiles of MgII line in the observed object.
The shape of the line has clearly changed during the $\sim 75$ days period between the first and the second exposure. The line intensity has changed by 9 \%. Thess facts allow for a systematic
monitoring of this source.

We now perform Monte Carlo simulations in order to optimize the project.

\section{Conclusions}

We propose that dust forms not only in the outer dusty/molecular torus, but also closer in, in the non-irradiated thin disk atmosphere. This dust is responsible for the formation of the failed wind which accounts for the formation of the Low Ionization Line part of the Broad Line Region.

This interpretation automatically explains why the size of the BLR scales with the square root of the monochromatic, instead of bolometric, luminosity. It also solves the problem of high turbulence in the region without signatures of the systematic outflow. Since the this mechanism of BLR formation seems universal, and there is no strong metalicity gradients in quasars up to high redshift, this also justifies the use Eq.~2.1 for distant quasars as a way to measure their luminosity distance and probe the dark energy distribution. 

{\underline{\it Acknowledgements}}. This work was supported in part by grants NN N203 387737, NN 203 580940 and Nr. 719/N-SALT/2010/0. 
All of the observations reported in this paper were obtained with the Southern African Large Telescope (SALT).

\end{document}